\documentclass[pra,twocolumn,showpacs]{revtex4}

\usepackage{color}
\usepackage{bm}
\usepackage{amsmath}
\usepackage{graphicx}
\usepackage{amssymb}
\usepackage{bm}
\usepackage{times}

\begin{document}

\title{Generating coherent state of entangled spins}
\author{Hongyi Yu} \author{Yu Luo} \author{Wang Yao}
\thanks{wangyao@hkucc.hku.hk}
\affiliation{Department of Physics and Center of Theoretical and
  Computational Physics, The University of Hong Kong, Hong Kong,
  China}

\begin{abstract}
  A coherent state of many spins contains quantum entanglement which
  increases with a decrease in the collective spin value. We present a
  scheme to engineer this class of pure state based on incoherent spin
  pumping with a few collective raising/lowering operators. In a pumping
  scenario aimed for maximum entanglement, the steady-state of $N$
  pumped spin qubits realizes the ideal resource for the $1
  \rightarrow \frac{N}{2}$ quantum telecloning. We show how the scheme can be implemented in a realistic system of atomic spin qubits in optical lattice. Error analysis show
  that high fidelity state engineering is possible for $N\sim O(100)$
  spins in the presence of decoherence. The scheme can also prepare a resource
  state for the secret sharing protocol and for the construction of
  large scale Affleck-Kennedy-Lieb-Tasaki (AKLT) state.
\end{abstract}
\date{\today}


\maketitle

\section{Introduction}

Coherent state in quantum mechanics usually refers to a specific type of quantum states with minimum uncertainty. It was first discovered in the context of oscillator field and has found wide applications in quantum optics~\cite{Zhang_review,Glauber_1963a,Glauber_1963b,Sudarshan_1963}. A quantum harmonic oscillator in coherent state most closely resembles the behavior of a classical oscillator. The notion was later generalized to spin systems~\cite{Klauder_1963a,Perelomov_SCS,gilmore_geometry_1972,arecchi_atomic_1972}. In an ensemble of $N$ spin-$I$ particles, the term spin coherent state (or atomic coherent state) is used to denote the states where the collective spin
$\hat{\mathbf J} \equiv \sum_n \hat{\mathbf I}_n$ has the minimum uncertainty~\cite{arecchi_atomic_1972}.
Such states can be easily identified in the basis $|J, \mu, \vec{\lambda}  \rangle$ which are eigenstates of $\hat{J}^2$ and $\hat J_z$ with eigenvalues $J(J+1)$ and $\mu$ respectively, and $\vec{\lambda}$ denotes additional quantum numbers to provide a complete set of labels. The collective spin $J$ gets every value from $NI$ down to $0$ (or $\frac{1}{2}$) in integer steps, and for each collective spin value $J$, the magnetic quantum number $\mu=-J, -J+1 \dots, J$. One can easily show that the minimum uncertainty relation $\langle (\Delta \hat{J}_x)^2 \rangle \langle (\Delta \hat{J}_y)^2 \rangle = \frac{1}{4}\langle\hat J_z\rangle^2$ is satisfied for all extremal states $|J, \mu = - J, \vec{\lambda}  \rangle$ in this basis. Hence these states and their rotations generated by $\hat{\mathbf J}$ are the spin coherent states (SCS)~\cite{arecchi_atomic_1972}. Interestingly, two contrary characters coexist on these states: the most classical collective spin
behavior; and the fundamentally non-classical phenomenon of quantum
entanglement. For every $J < NI$, there is a degenerate set of $|J,-J,
\vec{\lambda} \rangle$ with identical collective properties and
distinct entanglements where the number of unentangled spins is upper
bounded by $\frac{J}{I}$~\cite{tth_generation_2010}.

Preparation of SCS has been possible only in limited cases. The $J=NI$
SCS, nondegenerate and unentangled, is obtained when all spins are
fully polarized. Most experimental studies of spin squeezing start on
this state. Schemes were also proposed to populate mixed state of
singlets ($J=0$ SCS) by collective pumping~\cite{yao_many-body_2011},
and to select out singlet by projective measurement in a scattering
model~\cite{ciccarello_physical_2010}. Engineering a pure-state SCS of
an arbitrary $J$ value is a challenge but of multi-fold
significance. It is the sufficient condition for initialization into a
decoherence free subsystem for robust quantum computation under strong
collective decoherence ~\cite{kempe_theory_2001}. SCS of $J \ll NI$
are resources of large scale entanglement with potential uses in
one-way quantum
computation~\cite{gottesman_demonstrating_1999,raussendorf_one-way_2001}. The
ability to access a SCS of entangled spins also opens up a new realm
for the study of the interrelation between collective spin behaviors
and quantum correlations in a spin
ensemble~\cite{sorensen_entanglement_2001,guehne_entanglement_2009}.

In this paper, we propose control schemes for engineering pure-state
SCS of an arbitrarily specified collective spin value in a general spin ensemble. The schemes are based on incoherent
spin pumping of the $N$ target spins and a set of ancilla spins by a
few (e.g. three) collective raising/lowering operators. The desired
pure state is obtained with an $N$-independent probability by a single
projective measurement on the steady state of the pumping, and the
success rate approaches $100\%$ with $O(10)$ cycles of pump plus
measure. In a simplified pumping scenario aimed for maximum
entanglement, the steady state of $N$ pumped spin qubits (without
measurement projection) realizes the ideal resource for $1 \rightarrow
\frac{N}{2}$ optimal quantum
telecloning~\cite{murao_quantum_1999}. We show how the scheme can be implemented in the realistic system of atomic spin qubits trapped in optical lattice, where the collective spin pumping is realized by Stokes or anti-Stokes light scattering.
Error analyses show that high fidelity state engineering is possible for up to $N\sim O(100)$ atomic spin qubits in the presence of control errors and decoherence. This is a concrete example of
using simple and robust irreversible dynamics to prepare a desired
complicated quantum
state~\cite{clark,schneider,diehl,verstraete_stateprep,cho,Kastoryano}. The scheme can also prepare resource state for the secret sharing
protocol~\cite{4qubitsinglet}, and for efficient construction of large
scale AKLT state with applications in one-way quantum
computation~\cite{affleck_rigorous_1987,kaltenbaek_optical_2010,darmawan_optical_2010}.


\section{General scheme} 

Key to the state engineering approach by irreversible dynamics is to
design the dissipative controls under which the system saturates to
the desired state vectors. We utilize here the spin pumping process
which drives a spin system towards a mixture of all singlets connected
to the initial state by the pumping
operators~\cite{yao_many-body_2011}.  If a target spin ensemble is in
singlet with a spin-$J$ ancilla, its collective spin value must also
equal to $J$. With a proper constraint from conserved quantum numbers,
the singlet can be unique from which the desired pure-state SCS of the
target spins can be obtained.

The target spins are divided into two subgroups with collective spin
$\hat{\mathbf j}_A$ and $\hat{\mathbf j}_B$ respectively, and the
collective spin of $2J$ spin-$\frac{1}{2}$ ancillas is denoted
by $\hat{\mathbf j}_\beta$.  When an inhomogeneous collective operator
of the form $\hat{J}^+_i=c_A\hat{j}_{A}^{+}+c_B\hat{j}_{B}^{+}
+c_{\beta}\hat{j}_{\beta}^{+}$ acts on a SCS, the final state can be generally written as
\begin{equation}
  \hat J_i^+|J_T,-J_T,\vec\lambda\rangle=\sum_{\Delta,
    \vec\lambda'}\chi^{J_T+\Delta,\vec\lambda'}_{J_T,\vec\lambda}|J_T+\Delta,-J_T+1,\vec\lambda'\rangle.
\end{equation}
where the first two quantum numbers in the kets denote the total spin
and the $z$-component of $\hat{\mathbf J}_T=\hat{\mathbf j}_{A}
+\hat{\mathbf j}_{B}+\hat{\mathbf j}_{\beta}$ respectively.
Calculation of the coefficients $\chi$ is straightforward by expanding the collective spin states in terms of common eigenstates of $\hat{j}^2_{A}$, $\hat{j}^z_{A}$, $\hat{j}^2_{B}$, $\hat{j}^z_{B}$, $\hat{j}^2_{\beta}$ and $\hat{j}^z_{\beta}$. We find that only the $\Delta=0, \pm1$ transitions are allowed~\cite{yao_many-body_2011}, and the ratio between transition rates $\Delta =\pm1$ is
\begin{equation}
\left|\chi^{J_T,\vec\lambda'}_{J_T+1,\vec\lambda}\right|^2 =(J_T+1)(2J_T+1) \left|\chi^{J_T+1,\vec\lambda}_{J_T,\vec\lambda'}\right|^2. \label{ratio}
\end{equation}
Consider the incoherent strong pump by $\hat{J}_T^-$ which results in a mixture of SCS of
$\hat{\mathbf J}_T$, and the weak pump by the inhomogeneous operator
$\hat{J}^+_i$ which then causes transitions between these SCS with the
effective rate $\propto|\chi|^2$ and the selection rule $\Delta=0,
\pm1$. From Eq.~(\ref{ratio}), we can see the $\Delta=-1$ transition is much faster than the
$\Delta=1$ one between any such pair of states. Thus, the pump will
saturate the target and ancilla spins to singlets of $\hat{\mathbf
  J}_T$ where $J_T$ is minimized. With the target and ancilla spins initialized on the fully
polarized state, the quantum numbers $j_A=N_A I$, $j_B=N_B I$ and
$j_{\beta}=J$ are all conserved by the pump operators. Only one singlet
exists under this constraint: 
\begin{equation}
|S_{AB\beta}\rangle \equiv
\sum_{\mu=-J}^J\left(-\right)^{J-\mu}\left|J,\mu,j_{A},j_{B}\right\rangle
_{AB}\otimes\left|J,-\mu\right\rangle_{\beta},
\end{equation}
where $\left|J,\mu,j_{A},j_{B}\right\rangle_{AB}$ denotes eigenstates of
$(\hat{\mathbf j}_{A}+\hat{\mathbf j}_{B})^2$ and $\hat{j}^z_{A}+\hat{j}^z_{B}$ with eigenvalues $J(J+1)$ and $\mu$ respectively.

\begin{figure}[t]
\includegraphics[bb=15bp 46bp 570bp 432bp,clip,width=8cm]{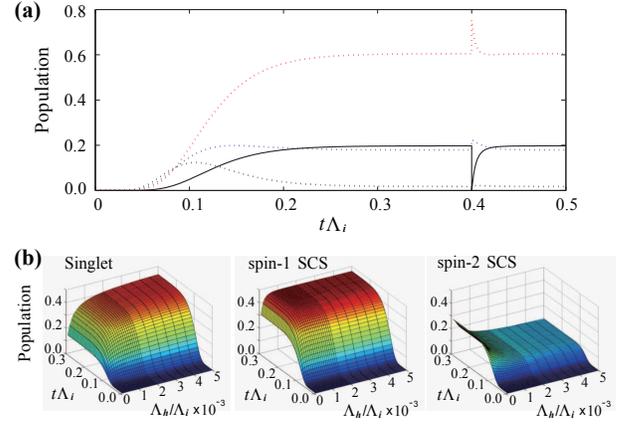}
\caption{ (a) Simulation of spin pump and repump by the collective
  operators $\hat{J}_T^-$, $\hat{J}_1^+$ and $\hat{J}_2^+$ (see
  text). The target and the ancilla spins are in the fully polarized
  state at $t=0$. Solid curve: population on the singlet
  $|S_{AB\beta}\rangle$. Dotted curves: populations in the subspace of
  $J=1$ (red), $J=2$ (blue), and $J=3$ (black) respectively. We assume
  the singlet is projected out at $t=0.4\Lambda_{i}^{-1}$, and hence
  the curve after is the repump dynamics.  (b) Simulation of the
  simplified scheme for engineering the telecloning resource for
  $N=40$ qubits. The populations on the singlet
  $|0,0,\frac{N}{4},\frac{N}{4} \rangle $, the spin-1 SCS
  $|1,-1,\frac{N}{4},\frac{N}{4} \rangle $, and the spin-2 SCS
  $|2,-2,\frac{N}{4},\frac{N}{4}\rangle$ are shown as functions of
  time using various pumping rates.}
\label{fig1} 
\end{figure}

Fig.~\ref{fig1}(a) presents a simulation of the spin pump using the
Lindblad master equation
$\dot{\rho}=-\frac{1}{2}\sum_{m=0}^{2}(\hat{L}_{m}^{\dagger}
\hat{L}_{m}\rho
+\rho\hat{L}_{m}^{\dagger}\hat{L}_{m}-2\hat{L}_{m}\rho\hat{L}_{m}^{\dagger})$,
where $\hat{L}_{0} \equiv \sqrt{\Lambda_h}\hat{J}_T^{-} $ and
$\hat{L}_{m} \equiv \sqrt{\Lambda_i}\hat{J}_{m}^{+} $ for
$m=1,2$. Here we have chosen the inhomogeneous raising operators
\begin{equation}
  \hat{J}_{1}^{+}=\mathrm e^{\frac{2}{3}\pi i}\hat{j}_{A}^{+}+\mathrm e^{\frac{4}{3}\pi i}\hat{j}_{B}^{+}+\hat{j}_{\beta}^{+},  ~~
  \hat{J}_{2}^{+}=\mathrm e^{\frac{4}{3}\pi i}\hat{j}_{A}^{+}+\mathrm e^{\frac{8}{3}\pi i}\hat{j}_{B}^{+}+\hat{j}_{\beta}^{+}, \notag
\end{equation}
while other choices of coefficients $c_{A,B,\beta}$ lead to similar
results. For the simulated example, we set the spin pump rates
$\Lambda_h/\Lambda_i=5000$, and $j_A =j_B =j_{\beta}=5$. After a pump
time $t_p = 0.2\Lambda_{i}^{-1}$, $|S_{AB\beta}\rangle$ is occupied by
a population $P\left(0\right)\sim20\%$. For general values of $j_A$,
$j_B$ and $j_{\beta}$, we require 
\begin{equation}
  \Lambda_h\langle\hat J_T^+\hat J_T^-\rangle\gg\Lambda_i\langle\hat J_i^-\hat J_i^+\rangle, \label{ratio2}
\end{equation}
which ensures the lowering operator $\hat J_T^-$ to be applied much more frequently than the raising operator $\hat J_i^+$.
The largest possible value of $\langle\hat J_i^-\hat
J_i^+\rangle$ is
$\sim(j_A+j_B+j_{\beta})^2$ while the smallest possible value of $\langle\hat J_T^+\hat J_T^-\rangle$ is $\sim 1$. Thus $ \Lambda_h/ \Lambda_i \gg (j_A+j_B+j_{\beta})^2$ is sufficient to ensure the condition in (\ref{ratio2}). The steady-state
population on $|S_{AB\beta}\rangle $ is given by $[\sum_k g(k) ]^{-1}
= 20 \%$ where $g (k) \equiv (2k+1)
\prod_{i=0}^{k-1}\left(2i^2+3i+1\right)^{-1}$~\cite{yao_many-body_2011}. The
timescale to reach steady-state is $\sim (j_A+j_B+j_{\beta})^{-1}
\Lambda_i^{-1} $. The singlet can be selected out by projective
measurement of $\hat{J}_T^{z}$ or $\hat{J}_T^{2}$. If the measurement
outcome is not singlet, the spins can be repumped to the steady state
in a much shorter timescale $\sim (j_A+j_B+j_{\beta})^{-2}
\Lambda_i^{-1}$ [Fig.~\ref{fig1}(a)]. The probability of NOT obtaining
$|S_{AB\beta}\rangle$ is reduced to $0.1\%$ after 30 cycles of measure
plus repump.

From the singlet $|S_{AB\beta}\rangle $, further pumping by the target spin operator $\hat{j}_{A}^{-}+\hat{j}_{B}^{-}$
bring the target spins to the desired SCS $\left|J, \mu=-J,j_A=N_A I, j_B=N_B I
\right\rangle$. Here the collective spin value $J$ of the target spins is controlled
by the number of the ancilla spins involved. Different choices of
$N_A$ and $N_B$ realize distinct pure SCS of the same collective spin
value, which are fully symmetric under permutation of spins within
subgroup A (or B). A and B can also be initialized with any $j_A<N_AI$
and $j_B< N_BI$ by applying the scheme first to the
subgroups. Concatenation of the scheme can therefore realize pure SCS
with more general permutation symmetries.

SCS of the smallest collective spin values are most desirable as a
resource of entanglement. We consider the $J=0$ scenario of the
above scheme (i.e.~no ancilla spins) which uses two pump operators:
the homogeneous $\hat{j}_{A}^{-}+\hat{j}_{B}^{-}$ and the
inhomogeneous $\hat{j}_{A}^{+}-\hat{j}_{B}^{+}$, where A and B each
contain $\frac{N}{2}$ target spins. For $\Lambda_{h}/\Lambda_{i}\gg
N^{2}$, we find the steady state of the pumping $\rho = \sum_J P(J)
|J, -J, \frac{N}{2}I, \frac{N}{2}I \rangle \langle J, -J,
\frac{N}{2}I, \frac{N}{2}I|$ where $P(J)=(2J^2+3J+1)P(J+1)$. This
steady state is reached with a pump time $t_p \approx \frac{3+ \ln
  NI}{2NI} \Lambda_i^{-1}$ by our numerical estimation, and is largely
a mixture of the singlet $|0,0, \frac{N}{2}I, \frac{N}{2}I\rangle$,
the spin-1 SCS $|1,-1,\frac{N}{2}I,\frac{N}{2}I \rangle $, and the
spin-2 SCS $|2,-2,\frac{N}{2}I,\frac{N}{2}I\rangle$, with some residue
population of $0.5\%$ on the spin-3 SCS
$|3,-3,\frac{N}{2}I,\frac{N}{2}I\rangle$. These states can be
distinguished in a non-demolition way by measuring
$\hat{j}_{A}^{z}+\hat{j}_{B}^{z}$. A single cycle of pump plus measure
thus ends up with one of these pure states which all have large scale
entanglement. Fig. \ref{fig1}(b) shows simulation of this spin pumping
for a cluster of $40$ spin qubits.

\begin{figure}[t]
\includegraphics[bb=15bp 55bp 365bp 320bp,clip,width=7cm]{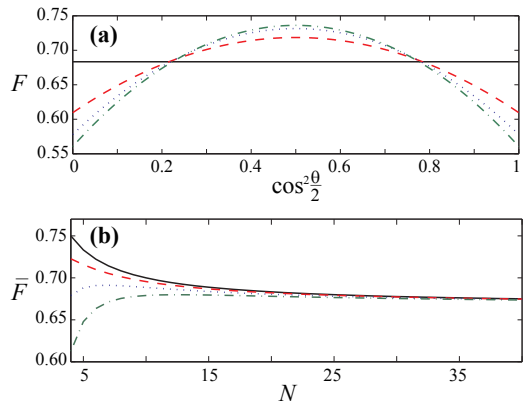}
\caption{ (a) Fidelity of $1 \rightarrow 20$ telecloning of a state
  $\cos\frac{\theta}{2}|0\rangle + \sin\frac{\theta}{2} e^{i \phi} |1
  \rangle$ using the above singlet (solid), spin-1 SCS (dashed),
  spin-2 SCS (dotted), and the spin-3 SCS
  $|3,-3,\frac{N}{4},\frac{N}{4} \rangle$ (dot-dashed) as the resource
  respectively. (b) The fidelity of $1 \rightarrow \frac{N}{2}$
  telecloning (averaged over the Bloch sphere) using the above
  resource states respectively. }
\label{fig2}
\end{figure}

The singlet $|0,0,\frac{N}{4},\frac{N}{4}\rangle $ of $N$ qubits turns
out to be the ideal resource for universal optimal quantum
telecloning~\cite{murao_quantum_1999}. If Alice holds subgroup A and
each of her $\frac{N}{2}$ associates holds a qubit in subgroup B,
Alice can transmit identical copies of her unknown state
$\cos\frac{\theta}{2} |0\rangle + \sin \frac{\theta}{2} e^{i \phi}
|1\rangle$ with a fidelity of $F_0=\frac{2N+2}{3N}$ to the
$\frac{N}{2}$ associates using local operations and classical
communications (LOCC)~\cite{murao_quantum_1999}. By a single cycle of
pump plus measure, the success rate to obtain this state is $\sim 46\%
$, which is a substantial improvement over the existing scheme where
the success rate is
$\frac{2}{2+N}$~\cite{ciccarello_physical_2010}. Most remarkably, all
alternative outcomes by our scheme,
i.e. $|J,-J,\frac{N}{4},\frac{N}{4} \rangle $ with a finite but small
$J$, can also be used as quantum telecloning resource under the same
LOCC. Following the same procedure of Ref.~\cite{murao_quantum_1999}
but replacing $|0,0,\frac{N}{4},\frac{N}{4}\rangle$ with
$|J,-J,\frac{N}{4},\frac{N}{4}\rangle$, we obtain the telecloning
fidelity which is then a function of $\theta$ [Fig.~\ref{fig2}(a)], and it reaches the maximum
value on the equator of the Bloch sphere.
\begin{equation}
  F_J^{\rm max} = \frac{(3J+4)N^2+4(J+1)N-4J(J+1)(J+2)}{2(2J+3)N^2}\nonumber\\
\end{equation}
Since $F_J^{\rm max} > F_0$, better telecloning fidelity can be achieved with these finite $J$ SCS in the presence of partial information (i.e. the range of the $\theta$ value). For $N \gg J$, the telecloning fidelity averaged over the entire Bloch sphere
approaches $F_0$ [Fig.~\ref{fig2}(b)]. Thus, the mixed steady state of
the spin pumping can be used as an equally efficient telecloning
resource as the ideal singlet.

\begin{figure*}[t]
\includegraphics[bb=17bp 76bp 556bp 472bp,clip,width=12cm]{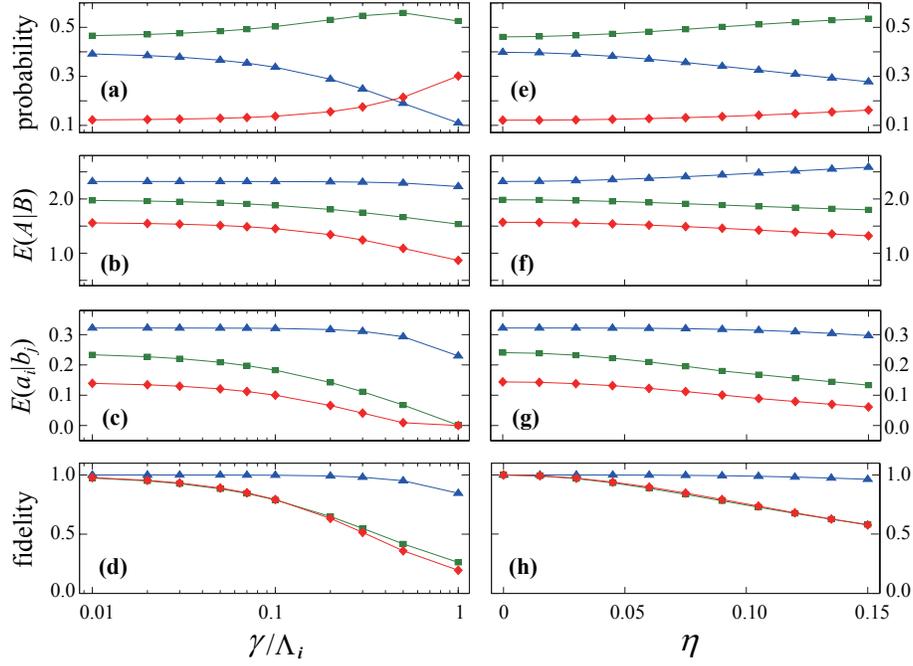}
\caption{Merits of state engineering in presence of spin decoherence
  (a-d) and system parameter errors (e-h) for engineering the
  telecloning resource for $N=8$
  qubits. $\Lambda_h/\Lambda_i=100$. (a) and (e): Probability of
  obtaining singlet (blue triangle), spin-1 SCS (green square) and
  spin-2 SCS (red diamond) at $t_p = 0.75 \Lambda_i^{-1}$. (b) and
  (f): Bipartite entanglement $E(A|B)$ in the obtained states measured
  with the logarithmic negativity.  (c) and (g): Entanglement between
  one qubit in A and another in B. (d) and (h): Fidelity of the
  obtained states with the target states
  $|0,0,\frac{N}{4},\frac{N}{4}\rangle$,
  $|1,-1,\frac{N}{4},\frac{N}{4}\rangle$ and
  $|2,-2,\frac{N}{4},\frac{N}{4}\rangle$.}
\label{fig3} 
\end{figure*}

A major cause of error for the state engineering is local spin
decoherence process. If each spin loses its phase coherence with a rate $\gamma$, a total leakage of $\sim N \gamma t_p \sim
\gamma/\Lambda_i$ out of the desired subspace is accumulated in the entire duration $t_p \sim
\frac{1}{N} \Lambda_i^{-1}$ of the state preparation. High fidelity
state engineering thus requires: $\gamma \ll \Lambda_i$. This is
confirmed by numerical simulation for a cluster of $N=8$ spin qubits where we have added pure dephasing processes described by Lindblad operators $\sqrt{2 \gamma} \hat{I}^z_n$ for all spins
[Fig.~\ref{fig3}(a-d)]. We also studied the effect of errors from
system parameters. For the simulation presented in
Fig.~\ref{fig3}(e-h), spins are pumped instead by
$\hat{\Xi}_{A}^{-}+\hat{\Xi}_{B}^{-}$ and
$\hat{\Xi}_{A}^{+}-\hat{\Xi}_{B}^{+}$ where
$\hat{\Xi}^{\pm}\equiv\sum_{n}(1+\eta_{n})\hat{I}_{n}^{\pm}$,
$\eta_{n}$ being a random error between $\eta$ and $-\eta$. The figure
of merit is reasonably good when the error amplitude $\eta <
10\%$. Moreover, by pumping with the operators
$U(\hat{j}_{A}^{-}+\hat{j}_{B}^{-})U^{\dagger}$ and
$U(\hat{j}_{A}^{+}-\hat{j}_{B}^{+})U^{\dagger}$ where $U \equiv
\prod_n \exp (i \theta_n \hat{I}_n^z)$, the state
$U\left|J,-J,\frac{N}{2}I,\frac{N}{2}I\right\rangle $ is obtained
instead of $\left|J,-J,\frac{N}{2}I,\frac{N}{2}I\right\rangle
$. Namely systematic phase errors in the collective pumping operators
do not affect the entanglement, and single spin rotations about the
$z$-axis can be deliberately encoded in the pumping.

\begin{figure}[t]
\includegraphics[bb=75bp 550bp 485bp 750bp,clip,width=8cm]{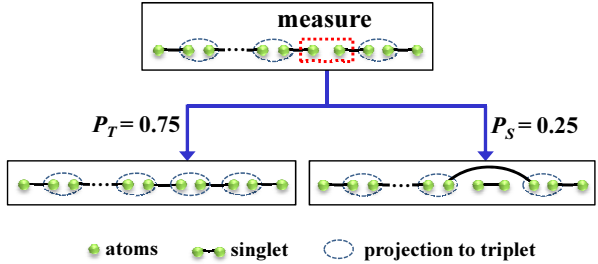}
 \caption{
   Increasing the length of a AKLT chain with the resource of
   4-qubit-AKLT-state. By measure the parity of the two atoms, the
   n-qubit AKLT chain either becomes a (n+4)-qubit (lower left) or a
   (n+2)-qubit AKLT chain (lower right). See text.}
\label{fig4} 
\end{figure}

When A and B each contains 2 qubits, the resultant singlet $|0,0, 1, 1
\rangle$ by our scheme is the 4-qubit AKLT state $P_{23}
|S\rangle_{12} |S\rangle_{34}$, where $|S\rangle_{ij}$ stands for the
singlet of qubit $i$ and $j$, and $P_{jk}$ is the projection operator
to the triplet subspace for qubit $j$ and
$k$~\cite{affleck_rigorous_1987}. Its optical analog has been used to
demonstrate four-party secret sharing~\cite{4qubitsinglet}, and
measurement based single qubit
rotation~\cite{kaltenbaek_optical_2010}.  This state is also an
efficient element to construct a large scale AKLT state as
schematically illustrated in Fig.~\ref{fig4}. Consider two 4-qubit
clusters in $P_{23} |S\rangle_{12} |S\rangle_{34}$ and $ P_{67}
|S\rangle_{56} |S\rangle_{78} $ respectively, by measuring the parity
of atom pair $\{4,5\}$, the spin configuration of this pair will be projected to either the singlet or the triplet subspace~\cite{Rey_paritymeasure_2007}. With $75\%$
probability, the measurement outcome is triplet and an $8$-qubit AKLT
chain
$P_{23}P_{45}P_{67}|S\rangle_{12}|S\rangle_{34}|S\rangle_{56}|S\rangle_{78}$
is obtained. The rest $25\%$ probability will give $P_{23} P_{67}
|S\rangle_{12} |S\rangle_{36} |S\rangle_{78} \otimes |S\rangle_{45}$
where a 6-qubit AKLT state is obtained. With our scheme as an efficient source of 4-qubit AKLT states, a long AKLT chain can thus be
constructed.

\section{Application to atoms in optical lattice} 
\begin{figure}[t]
\includegraphics[bb=90bp 360bp 460bp 810bp,clip,width=6cm]{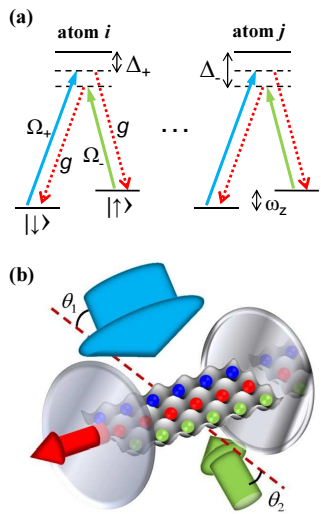}
\caption{(a) The level structure of the atoms. Red arrow represents
  the cavity field, and blue and green arrows denote two laser
  fields. (b) Collective spin pumping of atoms realized by cavity
  assisted Raman process. Atoms are trapped in optical lattice and
  loaded into a Fabry-Perot cavity. The inhomogeneous coefficients are
  controlled by laser emission angles $\theta_1$ and $\theta_2$.}
\label{fig5} 
\end{figure}

Here we apply our scheme to cold atoms of a typical $\Lambda$-level structure which are trapped in optical lattice. 
This system has been widely explored in various schemes for quantum
information processing. The two lower energy levels are used to represent the spin
qubit (Fig.~\ref{fig5}(a)). The atoms trapped in an optical lattice
can be first initialized to the $|\downarrow \rangle$ state and loaded
into a Fabry-Perot cavity (Fig.~\ref{fig5}(b)). For simplicity, we
assume the optical lattice constants in both directions equal to the
wavelength of the cavity mode and all atoms are at the peak of the
cavity field. Driven by two lasers of frequency $\omega_c\pm\omega_z$
with $\omega_c$ being the cavity resonance and $\omega_z$ the spin
splitting, the cavity-assisted Raman process lowers/raises the spin
state. The Raman processes in the large detuning regime
($g^2,\Omega_\pm^2\ll\Delta^2$) can be described by the effective
Hamiltonian
\begin{align}
  \hat H_\pm=\frac{g\Omega_\pm }{2\Delta_\pm}\hat J^\pm(\mathbf k)\hat
  a_{c}^\dag+\textrm{h.c.},
\end{align}
where $\Omega_\pm$ is the Rabi frequency of the two pumping lasers, $g$
the atom-cavity coupling and $\Delta_\pm$ the detuning. $\hat
a_{c}^\dag$ creates a cavity photon and $\hat J^\pm(\mathbf k) \equiv
\sum_j\mathrm e^{-i\mathbf k\cdot\mathbf r_j}\hat\sigma_j^\pm$
realizes various collective spin raising/lowering operations by
controlling the laser wavevector $\mathbf{k}$.  For the setup shown in
Fig.~\ref{fig5}(b) where the pump lasers are perpendicular to the
cavity axis, `blue' laser with $\theta_1=\frac{\pi}{2}$ realizes the
homogeneous operator $\hat{J}_T^{-}$, and `green' laser with
$\cos\theta_{2}=\frac{1}{3}$ (or $\frac{2}{3}$) realizes the
inhomogeneous collective operator $\hat{J}_{1}^{+}$ (or
$\hat{J}_{2}^{+}$) where subgroup A, B and ancilla are represented by
blue, red and green spheres respectively. 

Projective measurement for selecting out the singlet state can be realized in the same setup. Applying a `blue' and a `green' laser with both $\theta=\frac{\pi}{2}$ and comparable Rabi frequency $\Omega_-\sim\Omega_+$ realizes the
homogeneous raising and lowering operators $\hat{J}_T^\pm$ on the spin qubits. If the system is in finite $J$ state, then Raman scatterings are allowed and we shall observe continuous cavity photon emission when $\hat{J}_T^+$ and $\hat{J}_T^-$ pump the spins. When the system is in singlet, Raman scattering is forbidden since both $\hat{J}_T^+$ and $\hat{J}_T^-$ annihilate the state and there will be no cavity photon emission.

The Raman scattering rate by a single atom is $\Lambda_{h/i}= P\Gamma
\Omega_{\pm}^2 / \Delta_{\pm}^2$ with $\Gamma$ being the spontaneous
emission rate of atomic excited state in vacuum and $P$ the cavity
induced enhancement factor (Purcell factor).  Consider the Cs atom
($\frac{\Gamma}{2 \pi} =2.6$ MHz) in a typical Fabry-Perot cavity with
mode volume of $10^4~\mu$m$^3$and quality factor $1.7 \times 10^7$,
which correspond to $P \approx 80$, $\frac{g}{2 \pi} \approx 45$ MHz
and cavity decay rate $\frac{\kappa}{2 \pi} \approx 20$
MHz~\cite{hood_atom-cavity_2000}. $\Lambda \approx 15$ MHz can then be
achieved with $\frac{\Delta}{2 \pi} \approx 150$ MHz and
$\frac{\Omega}{2 \pi} \approx 40$
MHz~\cite{Mitsunaga_RabiFrequency_1996}. The collective Raman
scattering rates shall satisfy $\Lambda_i \langle \hat J_1^- \hat
J_1^+ \rangle \ll \Lambda_h \langle \hat J_T^+ \hat J_T^- \rangle <
\kappa$~\cite{yao_many-body_2011}, the last inequality is to ensure
the emission of cavity photon is spontaneous. For $N$ atoms in cavity,
the matrix element $\langle \hat J_T^+ \hat J_T^- \rangle$ ($\langle
\hat J_1^- \hat J_1^+ \rangle$ ) is $ \sim N$ ($\sim N$) in
the neighborhood of the polarized initial state and $\sim 1$ ($\sim
\frac{N^2}{4}$) in the neighborhood of the target singlet state. We
can thus use a $\Lambda_h \sim 15$ MHz. Correspondingly, $\Lambda_i$ shall be ramped down from an initial value of $\sim 1$ MHz to the steady state value
$\sim \frac{1}{N^2}$ MHz along with the spin pumping. Since atom in optical lattice can be of an
ultra-slow spin decoherence rate $\frac{\gamma}{2 \pi} \sim 1 - 25$
Hz~\cite{Rbdecohererate}, the condition $\gamma \ll \Lambda_i $ can be
satisfied for $N \sim O(100)$ qubits.

The authors acknowledge X. D. Xu for helpful comments. The work was supported by the Research Grant Council of Hong Kong under Grant No. 706711P.


\begin{thebibliography}{99}


\bibitem{Zhang_review}
W. M. Zhang,D. H. Feng and R. Gilmore, Rev. Mod. Phys. $\bf  62$, 867 (1990).


\bibitem{Glauber_1963a}
  J. R. Glauber, Phys. Rev. Lett. $\bf 10$, 84 (1963).

\bibitem{Glauber_1963b}
  J. R. Glauber, Phys. Rev. $\bf 130$, 2529 (1963).

\bibitem{Sudarshan_1963}
  E. C. G. Sudarshan, Phys. Rev. Lett. $\bf 10$, 277 (1963).

\bibitem{Klauder_1963a}
  J. R. Klauder, J. Math. Phys. $\bf 4$, 1055 (1963).

\bibitem{Perelomov_SCS}
  A. M. Perelomov, Commun. Math. Phys. $\bf26$, 222 (1972).

\bibitem{gilmore_geometry_1972}
  R. Gilmore, Ann. Phys. $\bf74$, 391 (1972).

\bibitem{arecchi_atomic_1972} 
  F. T. Arecchi, E. Courtens, R. Gilmore, and H. Thomas, Phys. Rev. A
  $\bf6$, 2211 (1972).

\bibitem{tth_generation_2010}
 G. T\'oth and M. W. Mitchell, New J. Phys. $\bf12$, 053007 (2010).

\bibitem{yao_many-body_2011}
  W. Yao, Phys. Rev. B $\bf 83$, 201308 (2011).

\bibitem{ciccarello_physical_2010}
  F. Ciccarello, M. Paternostro, S. Bose, D. E. Browne, G. M. Palma,
  and M. Zarcone, Phys. Rev. A $\bf 82$, 030302 (2010).


\bibitem{kempe_theory_2001}
  J. Kempe, D. Bacon, D. A. Lidar, and K. B. Whaley, Phys. Rev. A
  $\bf63$ 042307 (2001).

\bibitem{gottesman_demonstrating_1999}
  D. Gottesman and I. L. Chuang, Nature $\bf402$, 390 (1999).

\bibitem{raussendorf_one-way_2001}
  R. Raussendorf and H. J. Briegel, Phys. Rev. Lett. $\bf86$, 5188
  (2001).


\bibitem{sorensen_entanglement_2001}
  A. S. S{\o}rensen and K. M{\o}lmer, Phys. Rev. Lett. $\bf86$, 4431 (2001).

\bibitem{guehne_entanglement_2009}
  O. G\"uhne and G. T\'oth, Phys. Rep. $\bf474$, 1 (2009).

\bibitem{murao_quantum_1999}
  M. Murao, D. Jonathan, M. B. Plenio, and V. Vedral, Phys. Rev. A
  $\bf59$, 156 (1999).

\bibitem{diehl}
S. Diehl, A. Micheli, A. Kantian, B. Kraus, H. P. Buchler, and  P. Zoller, Nat. Phys. $\bf 4$, 878 (2008).

\bibitem{verstraete_stateprep}
F. Verstraete, M. M. Wolf, and  J. I. Cirac, Nat. Phys. $\bf 5$, 633 (2009) 

\bibitem{clark}
S. Clark, A. Peng, M. Gu, and S. Parkins, Phys. Rev. Lett. $\bf 91$, 177901 (2003). 

\bibitem{schneider}
S. Schneider and G. J. Milburn, Phys. Rev. A $\bf 65$, 042107 (2002).

\bibitem{cho}
J. Cho, S. Bose, and M. S. Kim, Phys. Rev. Lett. $\bf 106$, 020504 (2011).

\bibitem{Kastoryano}
M. J. Kastoryano, F. Reiter, and A. S. S{\o}rensen, Phys. Rev. Lett. $\bf 106$, 090502 (2011).

\bibitem{4qubitsinglet}
S. Gaertner, C. Kurtsiefer, M. Bourennane, H. Weinfurter, Phys. Rev. Lett. $\bf98$, 020503 (2007).

\bibitem{affleck_rigorous_1987}
  I. Affleck, T. Kennedy, E. H. Lieb, and H. Tasaki,
  Phys. Rev. Lett. $\bf59$, 799 (1987).

\bibitem{kaltenbaek_optical_2010}
  R. Kaltenbaek, J. Lavoie, B. Zeng, S. D. Bartlett, and K. J. Resch,
  Nat. Phys. $\bf6$, 850 (2010).

\bibitem{darmawan_optical_2010}
  A. S. Darmawan and S. D. Bartlett, Phys. Rev. A $\bf82$, 012328
  (2010).

\bibitem{Rey_paritymeasure_2007}
  A. M. Rey, V. Gritsev, I. Bloch, E. Demler and M. D. Lukin,
  Phys. Rev. Lett. $\bf99$, 140601 (2007).











\bibitem{hood_atom-cavity_2000}
  C. J. Hood, T. W. Lynn, A. C. Doherty, A. S. Parkins, and
  H. J. Kimble, Science $\bf287$, 1447 (2000).

\bibitem{Mitsunaga_RabiFrequency_1996}
  M. Mitsunaga, T. Mukai, K. Watanabe, and T. Mukai,
  J. Opt. Soc. Am. B, $\bf13$, 2696 (1996).

\bibitem{Rbdecohererate}
  U. Schnorrberger, J. D. Thompson, S. Trotzky, R. Pugatch,
  N. Davidson, S. Kuhr, and I. Bloch, Phys. Rev. Lett. $\bf103$,
  033003 (2009).
  R. Zhao, Y. O. Dudin, S. D. Jenkins, C. J. Campbell,
  D. N. Matsukevich, T. A. B. Kennedy, and A. Kuzmich, Nat.
  Phys. $\bf5$, 100 (2008).

\end{thebibliography}
\end{document}